\documentclass[12pt]{article}
\usepackage{amsmath}
\usepackage{amsfonts}
\usepackage[OT1]{fontenc}
\usepackage[english, francais]{babel}
\usepackage[applemac]{inputenc}
\usepackage{mltex}
\usepackage{graphics}

\newcommand{\nc}{\newcommand}
\nc{\beq}{\begin{equation}}
\nc{\edq}{\end{equation}}
\nc{\beqn}{\begin{eqnarray}}
\nc{\edqn}{\end{eqnarray}}
\nc{\N}{{\rm I\mkern-4.0mu N}}
\nc{\R}{{\rm I\mkern-4.0mu R}}
\nc{\Z}{{\sf Z\mkern-6.5mu Z}}
\def\C{ {\rm C \kern -.15cm \vrule width.5pt \kern .12cm}}
\nc{\Id}{{\mathchoice {\rm 1\mskip-4mu l} {\rm 1\mskip-4mu l}
{\rm 1\mskip-4.5mu l} {\rm 1\mskip-5mu l}}}
\newtheorem{lemma}{Lemma}[section]
\newtheorem{theorem}[lemma]{Theorem}
\newtheorem{proposition}[lemma]{Proposition}

\newtheorem{remark}[lemma]{Remark}

\newtheorem{definition}[lemma]{Definition}

\def\sq{\hbox {\rlap{$\sqcap$}$\sqcup$}}
\begin{document}
\title {THE QUANTUM FIDELITY FOR THE TIME
-PERIODIC SINGULAR HARMONIC OSCILLATOR}

\author{\it {\bf Monique Combescure} \\
\it Laboratoire de Physique Th\'eorique,   CNRS - UMR 8627\\
\it Universit\'e de Paris XI, B\^atiment
210, F-91405 ORSAY Cedex, France \\
\it and \\
\it IPNL, B\^atiment Paul Dirac \\
\it 4 rue Enrico Fermi,Universit\'e Lyon-1 \\
\it  F.69622 VILLEURBANNE Cedex, France\\
\it email monique.combescure@ipnl.in2p3.fr}
\maketitle

\bigskip

\begin{abstract}
In this paper we perform an exact study of ``Quantum Fidelity'' (also called Loschmidt 
Echo) for the time-periodic quantum Harmonic Oscillator of Hamiltonian :

\beq
\hat H_{g}(t):=\frac{P^2}{2 }+ f(t)\frac{Q^2}{2}+\frac{g^2}{Q^2}
\edq
when compared with the quantum evolution induced by $\hat H_{0}(t)$ ($g=0$), in
 the case where $f$ is a $T$-periodic function and $g$ a real constant. The reference (initial) 
 state is taken to be an arbitrary ``generalized coherent state'' in the sense of Perelomov.
 We show that, starting with a quadratic decrease in time in the neighborhood of $t=0$,
 this quantum fidelity may recur to its initial value 1 at an infinite sequence of times {$t_{k}$
 }. We discuss the result when the classical motion induced by Hamiltonian $\hat H_{0}(t)$ is
 assumed to be stable versus unstable. A beautiful relationship between the quantum and the
 classical fidelity is also demonstrated.
\end{abstract}

\newpage
\section{Introduction} 

A growing interest has been devoted recently on the study of the so-called ``Quantum Fidelity''
\beq
F_{g}(t):= \langle U_{0}(t,0)\psi, U_{g}(t,0)\psi \rangle
\edq
for some reference state $\psi$ that we take as initial wavepacket, $U_{0}(t,0)$ being the
quantum unitary evolution operator induced by some Hamiltonian , and $U_{g}(t,0)$ the 
quantum
evolution for a perturbation of it, $g$ being the size of the perturbation. The long-time
behavior of $F_{g}(t)$ is of particular interest, and it has been studied for a large class
of (time independent) Hamiltonians in a more or less heuristic way, and this long-time
behavior has been suggested to depend sensitively on the {\bf regular} versus {\bf chaotic}
motion of the underlying classical motion (see the bibliography).\\

In most of these heuristic works (\cite{beca1}, \cite{becave1}, \cite{becave2}, 
\cite{ceto1}, \cite{ceto2}, \cite{cupawi}, \cite{cupaja}, \cite{cudapazu}, \cite
{emwellco}, \cite{fihe}, \cite{gillma}, \cite{jaadbe1}, \cite{jaadbe2}, \cite{japa1},
\cite{japa2}, \cite{pro}, \cite{prose}, \cite{prosezni1}, \cite{prosezni2}, \cite{prozni},
\cite{schlunk}, \cite{sitwobe}, \cite{vaco}, \cite{vahe1}, \cite{vahe2}, 
\cite{wang1},
\cite{wang2}, \cite{welltsa}, \cite{weemllco}, \cite{wico}, \cite{znipro}), it is
claimed that the Quantum Fidelity decays very fast to zero as time grows, when the underlying
classical (unperturbed) dynamics is generically chaotic.\\

Although the short time decay of the fidelity is rather well understood (\cite{wi}), the arguments
put forward in the above cited works are not entirely convincing, since they are either purely
numerical, or extrapolate the ``short time'' behavior to guess the (Gaussian or exponential)
decay at $\infty$.\\

In other approaches (\cite{per}, \cite{jaadbe2}, \cite{welltsa}), the case of integrable
or regular systems is considered as well, and seems to indicate the occurrence of an anomalous
power law decay. Moreover, in a case of nearly integrable system a {\bf recurrence} of
fidelity has been exhibited; in this case the Quantum Fidelity manifests recurrences very
close to the initial value 1 as time evolves.\\

Thus it is a very intriguing subject of high interest to know more about the complete time
behavior of the Quantum Fidelity when the underlying dynamics is chaotic versus regular. To
our knowledge, no rigorous approaches of this topics have been attempted yet. It would be
highly desirable to have a mathematically explicit description of the long time behavior of
the Quantum Fidelity, although it is a very difficult task.\\

This paper is a first attempt towards this rigorous understanding. It relies on a rather
simplistic class of Hamiltonians for which the perturbed as well as unperturbed quantum
dynamics can be explicitly solved in terms of the classical dynamics. Moreover the reference
quantum states are taken in a suitable large class of quantum states known as ``generalized
Coherent States'' in the sense of Perelomov. The results are rather surprising:\\
- for a large class of  reference states, the Quantum Fidelity {\bf never decreases
to 0}, but instead remains bounded from below by some constant.\\
-in the unstable case, the Quantum Fidelity either decays exponentially fast to some non-zero
constant, as $t \to \infty$, or manifest strong recurrences to 1.\\
-in the stable case the Quantum Fidelity always manifests strong recurrences to 1 as time evolves.\\

These facts are strongly related to the  underlying $SU(1,1)$
structure underlying the corresponding Hamiltonians, as can be seen from the work of Perelomov
 (\cite{pe}).\\

We also are able to show a strong relationship between ``Quantum and Classical Fidelity'' for
 this specific situation.\\
 
 To complete this Introduction, let us notice that a notion of Classical Fidelity that ``mimics''
 the Quantum Fidelity has been proposed in the literature (\cite{ec}, \cite{beca2}, 
 \cite{vepro}), where decay properties similar to those of the Quantum Fidelity appear. Thus
 it would be desirable to understand more deeply the relationships between the Classical and
 Quantum Fidelity on a firm mathematical basis. We shall pursue this investigation in future
 publications, notably in the semiclassical limit (see \cite{coro2}, \cite{coco}).

 \section{Calculus of the Quantum Fidelity}
 
 Let us consider the following operators in $\mathcal H = L^2(\mathbb R)$ with suitable
 domains (see \cite{pe}):
 
 \beq
 K_{0}= \frac{Q^2+P^2}{4}+\frac{g^2}{2Q^2}
 \edq
 \beq
 K_{\pm}= \frac{Q^2- P^2}{4}\mp i \frac{Q.P+P.Q}{4}-\frac{g^2}{2Q^2}
 \edq
 where $Q$ is the usual multiplication operator by $x$ and
 $$P:=-i \frac{d}{dx}$$
 $K_{0}$ and $K_{\pm}$ satisfy the usual commutation rules of $SU(1,1)$ algebra, namely
 
 \beq
 [K_{0}, K_{\pm}]= \pm K_{\pm}
 \edq
 
 \beq
 [K_{-}, K_{+}]= 2K_{0}
 \edq
 
 We define the ``generalized coherent states'' (squeezed states) as follows: given any
 $\beta \in \mathbb C$
 \beq
 S_{\beta}= \exp (\beta K_{+}-\bar \beta K_{-})
 \edq
 
 \beq
 \psi_{\beta}= S(\beta)\psi
 \edq
 
 $\psi$ being a normalized state in $\mathcal H$ such that
 
 \beq
 K_{-}\psi=0
 \edq
 
 \beq
 2K_{0}\psi = (\alpha + \frac{1}{2})\psi
 \edq
 with
 \beq
 \alpha := \frac{1}{2}+ \sqrt{\frac{1}{4}+2g^2}
 \edq
 
 We shall focus on the following cases $g=1, \ 
 g=\sqrt 3$, where $\psi$ has the following
 form:
 
 \beq
 \varphi(x) = c_{1}x^2 e^{-x^2/2}
\edq

\beq
\chi(x)= c_{2}x^3 e^{-x^2/2}
\edq

with
\beq
c_{1}= \sqrt {\frac{4}{3}}
\edq

\beq
c_{2}= \sqrt {\frac{8}{15}}
\edq
(omitting the factors in $\pi$).
\\

This makes $\varphi$ and $\chi$ to be (normalized to 1) finite linear combinations of 
Hermite functions.\\

It has been shown (see \cite{pe}) that for $\psi = \varphi, \ \psi_{\beta}$ has the
following form:

\beq
\psi_{\beta}= c_{1}x^2 e^{-2(u-\epsilon)}\exp \left(-\frac{5i\theta}{2}+
i\frac{\dot u x^2}{2}-\frac{1}{2}(u-\epsilon)-\frac{1}{2}(xe^{-(u-\epsilon)})^2
\right)
\edq
where the constants $u, \dot u, \theta, \epsilon$ are suitably determined from $\beta \in
\mathbb C$ , whereas for $\psi= \chi$, $\psi_{\beta}$ is:

\beq
\psi_{\beta}= c_{2}x^3 e^{-3(u-\epsilon)}\exp \left(-\frac{7i\theta}{2}+
i\frac{\dot u x^2}{2}-\frac{1}{2}(u-\epsilon)-\frac{1}{2}(xe^{-(u-\epsilon)})^2
\right)
\edq

Now the evolutions of $\psi_{\beta}$ with respect to $U_{1}(t,0)$ and $U_{\sqrt 3}(t,0)$
respectively, together with $U_{0}(t,0)$ are completely explicit.\\

Consider a complex solution of the classical equations of motion induced by Hamiltonian
$\hat H_{0}(t)$:
\beq
\ddot x + f x =0
\edq
and look for it in the form

\beq
x=e^{u+i\theta}
\edq
where the functions $t \mapsto u$ and $t \mapsto \theta$ are {\bf real}.
\\

We assume for $u(t)$ and $\theta(t)$ the following initial data:
$$u(0)=u_{0}$$
$$\dot u(0)= \dot u_{0}$$
$$\theta(0)= \theta_{0}$$
$$\dot \theta(0)= e^{-2(u_{0}-\epsilon)}$$
 Since $f$ is real, the wronskian of $x$ anf $\bar x$ is constant and equals
 \beq
 2ie^{2\epsilon} = 2i \dot \theta(t)e^{2u(t)}
 \edq
 
 This yields
 \beq
 \ddot x = [\ddot u + i \ddot \theta +(\dot u + i \dot \theta)^2]x =
  [\ddot u + \dot u^2-e^{-4(u-\epsilon)}]x = -fx
  \edq
  and therefore $u$ obeys the following differential equation
  \beq
  \ddot u+ \dot u^2 -e^{-4(u-\epsilon)}+f =0
  \edq
  
  We have the following result:\\

  \begin{lemma}
  Let $\hat H_{g}= \frac{P^2+Q^2}{2}+ \frac{g^2}{Q^2}$. 
  Then the quantum propagator for $\hat H_{g}(t)$ is of the following form:
  
  \beq
  U_{g}(t,0)= e^{i\dot uQ^2/2}e^{-i(u-\epsilon)(Q.P+P.Q)/2}e^{-i(\theta- \theta_{0}
  )\hat H_{g}}e^{i(u_{0}-\epsilon)(Q.P+P.Q)/2}e^{-i\dot u_{0}Q^2/2}
  \edq
  The same formula holds for the propagator $U_{0}(t,0)$ of $\hat H_{0}(t)$ with
   $\hat H_{g}$ replaced by $\hat H_{0}$.
     \end{lemma}
     
     \bigskip
     Proof: Let us denote
     \beq
     V(t):= e^{i\dot uQ^2/2}e^{-i(u-\epsilon)(Q.P+P.Q)/2}e^{-i\theta \hat H_{g}}
     \edq

We have:
\beq
i\frac{d}{dt}V(t)= \left( -\ddot u \frac{Q^2}{2}+ \frac{\dot u}{2}
e^{i\dot u Q^2/2}(Q.P+P.Q)e^{-i\dot u Q^2/2}\right)V(t)
\edq
$$+\left(\dot \theta e^{i\dot u Q^2/2}e^{i(u-\epsilon)(Q.P+P.Q)/2}\hat H_{g}
e^{-i(u-\epsilon)(Q.P+P.Q)/2}e^{-i\dot uQ^2/2}\right)V(t)$$

and since
\beq
e^{i\dot uQ^2/2}P e^{-i\dot uQ^2/2}= P-\dot u Q
\edq
we have:
\beq
e^{i\dot uQ^2/2}(P.Q+Q.P)e^{-i\dot uQ^2/2} = P.Q+Q.P-2i\dot u Q^2
\edq

Therefore the first line in the RHS of (25) is
\beq
\left((-\frac{\ddot u}{2}-\dot u^2)Q^2 + \frac{\dot u}{2}(Q.P+P.Q)\right)V(t)
\edq

Furthermore
\beq
e^{-i(u-\epsilon)(Q.P+P.Q)/2}\hat H_{g}e^{-i(u-\epsilon)(P.Q+Q.P)/2}=
 \left(\frac{P^2}{2}+\frac{g^2}{Q^2}\right)e^{2(u-\epsilon)}+ \frac{Q^2}{2}
 e^{-2(u-\epsilon)}
 \edq
 which implies that the second line in the RHS of (25) is
 \beq
 \left(\frac{1}{2}(P-\dot uQ)^2+ \frac{g^2}{Q^2}+ \frac{Q^2}{2}
 e^{-4(u-\epsilon)}\right)V(t)
\edq

where we have used that

$$\dot \theta= e^{-2(u-\epsilon)}$$

 Collecting the different terms, we get:
 \beq
 i\frac{d}{dt}V(t)= \left( \frac{P^2}{2}+ \frac{g^2}{Q^2}+ \frac{Q^2}{2}
 (-\ddot u -\dot u^2 +e^{-4(u-\epsilon)})\right)V(t)
 \edq
 $$= \left(\frac{P^2}{2}+ \frac{g^2}{Q^2}+f(t)\frac{Q^2}{2}\right)V(t)$$
  using (22).
  \sq\\
  
  We shall now consider the ``quantum fidelity'' in two particular cases $g=1$ and
   $g=\sqrt 3$, starting respectively with the initial states $\psi_{\beta,1}= 
   S(\beta)\varphi$, and $\psi_{\beta,2}= S(\beta)\chi$:
   
   \beq
   F_{1}(t)= \langle U_{0}(t,0)\psi_{\beta,1}, U_{g}(t,0)\psi_{\beta,1}\rangle
   \edq
   and similarly for $F_{2}(t)$ with $\psi_{\beta,1}$ replaced by $\psi_{\beta,2}$.
   
   \begin{theorem}
   (i) Let $g=1$ and $\psi_{\beta,1}= S(\beta)\varphi$. Then we have
   \beq
   F_{1}(t)= \frac{2}{3}+\frac{1}{3}e^{-2i(\theta(t)-\theta(0))}
   \edq
   (ii) Let $g= \sqrt 3$ and $\psi_{\beta,2}= S(\beta)\chi$. Then we have:
   \beq
   F_{2}(t)= \frac{2}{5}+\frac{3}{5}e^{-3i(\theta(t)-\theta(0))}
   \edq
   \end{theorem}
   
   \bigskip
   Proof: Since $x^2$  is expanded as
   $$x^2 = \frac{1}{4}H_{2}(x)+\frac{1}{2}H_{0}(x)$$
   it is clear that:
   
   \beq
   U_{0}(t,0)\psi_{\beta,1}= c_{1} \exp\left( \frac{i}{2}\dot u(t)x^2-\frac{1}{2}
   (u(t)-\epsilon)\right)
   \edq
   $$\times \left\{\frac{1}{4}e^{-5i\theta(t)/2}H_{2}(x e^{-(u(t)-\epsilon)})
   +\frac{1}{2}e^{-i\theta(t)/2 -2i\theta(0)}\right\}\exp \left( -\frac{1}{2}
   x^2 e^{-2(u(t)-\epsilon)}\right)$$
   
 and 
 \beq
 U_{g}(t)\psi_{\beta,1}= V(t)\varphi
 \edq
 $$= c_{1}\exp \left( -\frac{5}{2}i \theta(t)+\frac{i}{2}\dot u(t)x^2-
 \frac{1}{2}(x e^{-(u(t)-\epsilon)})^2\right)$$ 
 
 from which we deduce that
 
 \beq
 F_{1}(t)= c_{1}^2\int \left[x^2(x^2-\frac{1}{2})+\frac{e^{-2i(\theta(t)-\theta(0))}}
 {2}x^2\right]e^{-x^2}\ dx 
\edq
$$= \frac{4}{3}\left(\frac{1}{2}+\frac{1}{4}e^{-2i(\theta(t)-\theta(0))}\right)
= \frac{2}{3}+\frac{1}{3}e^{-2i(\theta(t)-\theta(0))}$$

(ii) follows from a very similar calculation using that

$$H_{3}(x)= 8x^3 -12 x$$
\sq\\

Let us now assume that $g$ has an arbitrary real value (not specifically of the form
$g=\sqrt {k(k+1)/2}$ for $k \in \mathbb N$ which gives rise to integer values of
 $\alpha = \frac{1}{2}+\sqrt {\frac{1}{4}+2g^2}$). Then the state $\psi:= 
 cx^{\alpha}e^{ -x^2/2}$ is no longer a finite linear combination of Hermite functions.
 It has nevertheless an infinite expansion on the basis of Hermite functions $\varphi_{n}$:
 
 $$\psi = \sum_{n=0}^{\infty}\lambda_{n}\varphi_{n}$$
 
 \noindent
 and one can establish the following general result about the corresponding ``quantum fidelity'':
 
 \begin{theorem}
 $$F_{g}(t)= \exp(-i\alpha(\theta(t)-\theta(0)))\sum_{n=0}^{\infty}\vert \lambda_{n}
 \vert^2 \exp(in(\theta(t)-\theta(0)))$$
 
 \noindent
 which, since $\sum_{n=0}^\infty \vert \lambda_{n}\vert^2=1$ , returns in 
 absolute value to 1 as long as $\ \theta(t)-\theta(0)= 0\ (\mbox{mod}\ 2\pi)$. If
 $\alpha = p/q$ is rational, then the quantum fidelity recurs exactly to 1 (not only in
 absolute value) provided that $\ \theta(t)-\theta(0)=0\ (\mbox{mod}\ 2q\pi)$.
 \end{theorem}
 
 Proof: Again according to ref. \cite{pe}, we have
 
 $$\psi_{\beta}:= S(\beta)\psi = e^{-i(\alpha+\frac{1}{2})\theta_{0}}D(u_{0})\psi$$
  where by $D(u)$ we denote the following unitary operator
  
  $$D(u):= e^{i\dot u Q^2/2} e^{-i(u-\epsilon)(Q.P+P.Q)/2}$$
  
  Then
  
  $$U_{0}(t,0)\psi_{\beta}= D(u_{t}) e^{-i(\theta_{t}-\theta_{0})\hat H_{0}}
  \sum_{n=0}^{\infty} \lambda_{n} e^{-i\theta_{0}(\alpha+\frac{1}{2})} 
  \varphi_{n}$$
  
  $$= D(u_{t})\sum_{n=0}^{\infty} \lambda_{n} e^{-i\theta_{t}(n+\frac{1}{2})
  + i\theta_{0}(n-\alpha)}\varphi_{n}$$
  
  whereas
  
  $$U_{g}(t,0)\psi_{\beta}= e^{-i\theta_{t}(\alpha+\frac{1}{2})} D(u_{t})\psi$$
  
  so that:
  
  $$\langle U_{0}(t,0)\psi_{\beta}, U_{g}(t,0)\psi_{\beta}\psi_{\beta}\rangle
  = \sum_{n,m}\bar \lambda_{n}\lambda_{m} e^{i\theta_{t}(n+\frac{1}{2})
  - i\theta_{0}(n- \alpha)-i\theta_{t}(\alpha +\frac{1}{2})}\langle \varphi_{n}
  , \varphi_{m}\rangle$$
  
  $$= \sum_{n=0}^{\infty} \vert \lambda_{n}\vert ^2 e^{i(\theta_{t}-\theta_{0})
  (n-\alpha)}$$
  \sq\\

  \section{Discussion of the Result}
  
  Since $f$ is a $T$-periodic function, Floquet analysis applies to equation (18) which is
   nothing but the well-known Hill's equation. Depending to the parameters characterizing
    the function $f$ (as for example $\gamma$ and $\delta$ in the case of Mathieu equation
     where $f(t)= \gamma + \delta \cos \omega t$), the solutions can be either {\bf stable}
     or {\bf unstable}. In all cases the quantum fidelity $F_{1}(t)\ \mbox{and}\ F_{2}(t)$
     are bounded from below by some positive constant in absolute value, and therefore never 
     decrease to zero.
\\
The phase $\theta(t)$ is determined by

\beq
\theta(t)- \theta(0) = e^{2\epsilon}\ \int_{0}^t e^{-2u(s)}\ ds
\edq

In the {\bf stable case}, the function $t \mapsto u(t)$ is $T$-periodic. Therefore $\theta(t)$
grows from $-\infty$ to $+\infty$ as time evolves from $-\infty$ to $+\infty$.
It follows that there exists an infinite sequence of times $\{t_{k}\}$ where $F_{1}(t)$ recurs
to its initial value 1, and an infinite squence $\{t'_{k}\}$ where $\vert F_{1}(t)\vert$ attains its
minimum 1/3. The same statement holds for $F_{2}(t)$ where 1/3 is replaced by 1/5.

\bigskip
\includegraphics{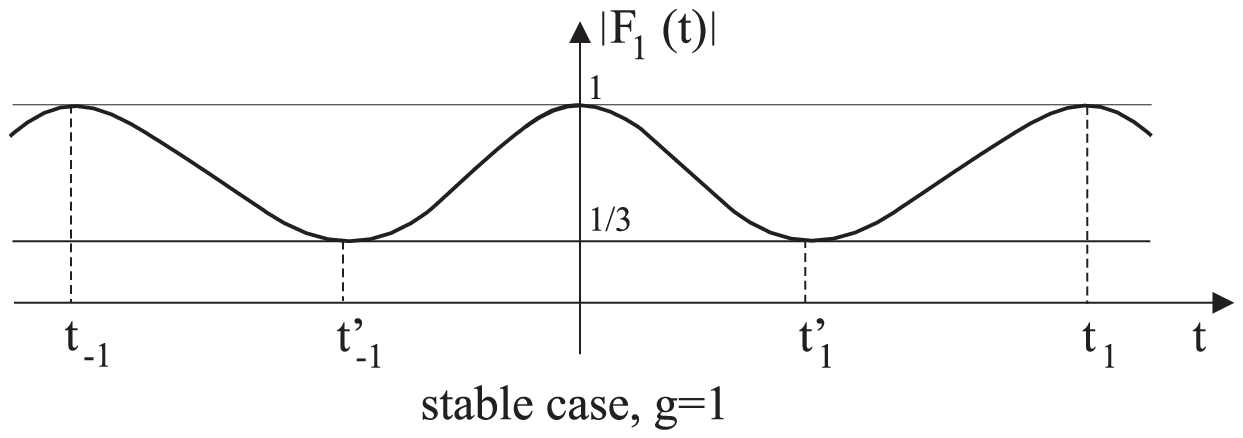}

\bigskip

In the {\bf unstable case} there is some positive Lyapunov exponent $\lambda$ and some real 
solution of Hill's equation such that

$$x(T)= e^{\lambda T}x(0)$$

$$\dot x(T)= e^{\lambda T}\dot x(0)$$

Here, since we deal with complex solutions $x(t)$ such that the Wronskian of $x$ and 
$\bar x$ is non zero, we deduce that $\vert x(t)\vert >0$. Moreover, depending on the
instability zone, $\vert x(t) \vert ^{-2}= e^{-2u(t)}$ can be integrable near $\pm \infty$
, or diverge at $\pm \infty$. These topics are clearly detailed in ref. \cite{mi}. In the first
 case the conclusion is that there exists two constants $\theta_{\pm}$ such that
 
 \beq
 \theta(t) \to \theta_{\pm}\ \mbox{as}\ t \to \pm \infty
 \edq
 
 This is the case for the Inverted Harmonic Oscillator ($f=-1$) that we describe in the last Section.
 \\
 Therefore the Quantum Fidelity in this case behaves generically as described in the following
  picture:
  
  \bigskip
\includegraphics{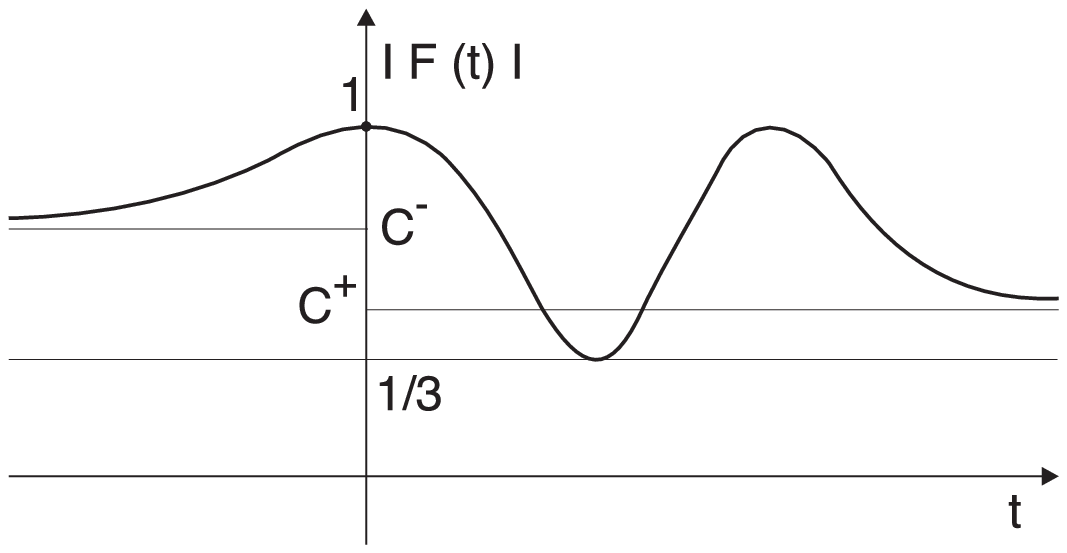}

\bigskip
If the instability zone in which the solution $x(t)$ lies is such that $\vert x(t)\vert^{-2}$ is
not integrable near $\pm \infty$, then the figure is typically similar to that of the stable
case, which shows infinite recurrences to 1 of the Quantum Fidelity.

\section{Link with the ``Classical Fidelity''}

We can call ``classical infidelity'' the discrepancy between two classical trajectories, along
their evolution, when they merge from the {\bf same initial phase-space point} at $t=0$.
 The classical fidelity is thus the possible  crossing of the two trajectories governed by 
 $H_{0}(t)$ and $H_{g}(t)$, in phase-space, as time evolves.\\
 From now on we assume that $\epsilon$ of Section 2 is determined by ($g>0$):
 
 \beq
 e^{2\epsilon}= g \sqrt 2
 \edq
 
 \begin{proposition}
 Let $x(t)$ be a {\bf real} trajectory for the time-periodic Hamiltonian $H_{0}(t)$, and
 $y(t)$ a {\bf real} trajectory for Hamiltonian $H_{g}(t)$. Assume in addition that they 
 satisfy $\ x(0)= y(0)\in \mathbb R,\qquad \dot x(0)= \dot y(0)\in \mathbb R$. Then
 we have:
 
 \beq
 \vert x(t)-y(t) \vert = \vert y(t)\vert (1- \cos \tilde \theta(t))
 \edq
 \noindent
 where $\tilde \theta(t) $ is defined as follows:
 
 \beq
 \tilde \theta(t):= g \sqrt 2 \int_{0}^t y(s)^{-2}\ ds \equiv \theta(t)-\theta(0)
 \edq 
 \end{proposition}
 
 \bigskip
 Proof: Let $z(t)$ be a complex solution of equation
 
 \beq
 \ddot z + f z =0
 \edq
 It can be written as $z(t)= e^{u+i\tilde \theta}$ as above, where $t \mapsto u\ \mbox
 {and}\ t \mapsto \tilde \theta$ are {\bf real} functions. 
 Assume that $x(t)$ is such that $x(0)>0$.
 Then it is easy to see that $x(t)= \Re z(t)= e^{u(t)}\cos \tilde \theta(t)$ is a solution of
 equation (43), and that the positive function $y(t):= e^{u(t)}$ is a solution of
 
 \beq
 \ddot y + f y -\frac{2g^2}{y^3}=0
 \edq
 which means that it defines a classical trajectory for $H_{g}(t)$.\\
 
 \noindent
 Namely we get from equ. (42) that 
 
 \beq
 \frac{d}{dt}\tilde \theta(t) = g \sqrt 2 e^{-2u(t)}
 \edq
 
 and therefore
 
 \beq
 \ddot x = (\ddot u + \dot u^2 -2g^2 e^{-4u})x = -f x
 \edq
 
 so that
 
 \beq
 (\ddot u + \dot u^2 -2g^2 e^{-4u}+f)e^{u}=0
 \edq
 
 which is nothing but equ. (44) noting that 
 \beq
 \ddot y = (\ddot u + \dot u^2)y
 \edq
 
 Moreover, they satisfy $\ x(0)= y(0),\qquad \dot x(0)= \dot y(0)$, and
 
 $$\vert x(t)- y(t)\vert = y(t) (1- \cos \tilde \theta(t))$$
 
 In the case where $x(0)<0$, we just take $x(t)= - \Re z(t)$ and $y(t)= -e^{u(t)}$, which 
 completes the result.
 \sq

 \bigskip
 {\bf Conclusion}: The classical infidelity vanishes for $\tilde \theta(t) = 2k \pi
 \ (k \in \mathbb Z)$, which precisely gives rise to recurrences to 1 of the ``quantum
 fidelity'' (Theorem 2.3). We expect this remarkable property to be true in more general 
 situations, in particular in the Semiclassical Regime (see \cite{coro2}).
 
 \bigskip
 By ``vanishing of the classical infidelity'' (thus ``classical fidelity'') we meant that given 
 any solution of the unperturbed dynamics, there exists a solution of the perturbed one that 
 coincides with it at the origin, and at any values of time solving the equation
 
 $$g \sqrt 2 \int_{0}^{t} ds\ y(s)^{-2}=0\ (\mbox{mod} \ 2\pi)$$
 
 One may ask whether this holds true for {\bf general} solution $x(t),\quad y(t) $ of equ.
 (43-44). This is answered in the following particular cases below.
 
 $\bullet\ ${\bf Particular case $f=1$}\\

 (The case $f = \omega^2$ could be treated as well.)
 \\
 When we apply the result above, we find that solutions $x(t)= A \cos t$ and $y =A$ 
 coincide for $t = 2k \pi, \ k \in \mathbb Z$, and that $y$ obeys equation (44) provided
  $A = (2g^2)^{1/4}.$  One shows that a more general result holds, involving 
  {\bf general solutions} of the equations under consideration.\\

  The general solution of equ. (44) is of the form (apart from the sign before the square root):
  
  \beq
  y(t)= \sqrt {\alpha + \beta \cos 2t + \gamma \sin 2t}
  \edq
  
  with $\alpha, \ \beta, \ \gamma$ related to each other by the relation
  
  \beq
  \alpha^2 -(\beta^2+ \gamma^2)= 2 g^2
  \edq
  
  It obeys the initial data
  
  $$y(0)= \sqrt {\alpha + \beta}$$
  
  $$\dot y(0)= \frac{\gamma}{\sqrt{\alpha + \beta}}$$
  
  and the conserved energy is simply $\alpha$.\\
  
  We take as real solution of equ. (43) (Harmonic Oscillator):
  
  \beq
  x(t)= \sqrt{\alpha+ \beta}\cos t + \frac{\gamma}{\sqrt {\alpha+ \beta}}\sin t
  \edq
  
  Which has the same initial data as $y(t)$. Both functions being $2\pi$- periodic, the generic 
  ``classical infidelity''  vanishes when $t= 2k \pi, \ k \in \mathbb Z$.\\

  $\bullet\ $ {\bf Particular case $f=-1$}\\
  
  (The case $f= -\omega^2$ could be trated as well)\\
  Any complex solution of the differential equation (Inverted Harmonic Oscillator)
  \beq
  \ddot z - z =0
  \edq
  
   can be
  written in the form
  
  \beq
  z(t)= (a+ib)e^t + (c+id)e^{-t}
  \edq
  
  where $a,\ b, \ c,\ d$ are real constants. We define:
  
  $$Z(t):= \vert z(t)\vert^2$$
  
  One can prove that $y(t)= \sqrt {Z(t)}$ obeys the differential equation
  
  \beq
  \ddot y - y -\frac{2g^2}{y^3}=0
  \edq
  
  with
  \beq
  g^2 = 2(ad-bc)^2>0
  \edq
  
  It is important to note that this implies $Z(t)>0, \quad \forall t$.\\
  
  Define $x(t):= y(t) \cos \theta(t)$, with $\theta(0)=0$. Then, clearly
  
  $$x(0)= y(0)$$
  
  $$\dot x(0)= \dot y(0)$$
  
  We want $x(t)$ to be a real solution of equ.(52). Since
  
  \beq
  \ddot x -x = \frac{1}{\sqrt Z} \left(\frac{\ddot Z}{2}- \frac
  {\dot Z^2}{4Z}-Z -\dot \theta^2 Z \right)\cos \theta - 
  \left( \frac{\dot \theta \dot Z}
  {\sqrt Z}+ \ddot \theta \sqrt Z \right)\sin \theta
  \edq
  
  and since $Z $ obeys:
  
  \beq
  \frac{\ddot Z}{2}-\frac{\dot Z^2}{4Z}-Z-\frac{2g^2}{Z}=0
  \edq
  
  the RHS of equ. (56) vanishes provided
  
  \beq
  \dot \theta^2 = 2g^2 Z^{-2}
  \edq
  
  Then $x(t)$ and $y(t)$ are two trajectories for Hamiltonians $ \frac{P^2-Q^2}{2}$
  and $\frac{P^2-Q^2}{2}+\frac{g^2}{Q^2}$ respectively, with the same initial
  data, and we have
  
  \beq
  \vert x(t)-y(t)\vert = y(t)(1-\cos \theta(t) )
  \edq
  
  Using the particular solution of equ. (52)
  
  $$z(t)= \frac{1+i}{2}e^t + \frac{1-i}{2}e^{-t}$$
  
  that can be rewritten as:
  
  $$z(t)= \sqrt {\cosh 2t}\exp (\frac{i}{2} \arcsin \tanh 2t)$$
  
  i.e. with $u(t)= \frac{1}{2}\log \cosh 2t$ (which satisfies $u(0)= \dot u(0) =0$), the
  formula (23) can be rewritten in the simple form
  
  \beq
  U_{g}(t,0)= \exp \left( \frac{i}{2}Q^2 \tanh 2t\right) \exp \left(\frac{i}{4}
  (Q.P+P.Q)\log \cosh 2t\right) \exp \left(-\frac{i}{2}\hat H_{g}
  \arcsin \tanh 2t\right) 
  \edq
  
  which holds true for $g=0$, and $g=1/\sqrt2$. For $g=0$, this is nothing but the 
  well-known Mehler's Formula.
  \\
  
  {\bf Study of $\theta(t)$}
  
  Recall that the fidelity $F_{g}(t)$ strongly depends on the reference state via the constants
  $a,\ b, \ c, \ d$
  
  \beq
  \theta(t)=  g \sqrt 2 \int_{0}^t \frac{ds}{(a^2+b^2)e^{2s}+(c^2+d^2)e^{-2s}
  +2(ac+bd)}
  \edq
  
  $$= 
  \left( \arctan \left(\frac{ac+bd +(a^2+b^2)e^{2t}}{\vert ad-bc\vert}
  \right)- \arctan\left( \frac{ac+bd+(a^2+b^2)}{\vert ad-bc\vert}\right)\right)
   $$
   
   This implies that $\theta(t) \to \theta_{\pm}\ \mbox{as}\ t \to \pm \infty$
    exponentially fast in the future and in the past. The calculus is especially simple in
     the particular case where we choose $a=d= g/ \sqrt {2},\ c=b=0$:
     
     \beq
     \theta(t)= \arctan(e^{2t})-\frac{\pi}{4}
     \edq
     
     No classical fidelity happens, and the square of the quantum fidelity in the case $g=1$
     is nothing but
     
     \beq
     \vert F_{1}(t)\vert^2 = \frac{5}{9}+ \frac{8}{9(e^{2t}+e^{-2t})}
     \sim \frac{5}{9}+ \frac{4}{9} e^{-2\vert t \vert}
     \edq
     
     as $t \to \pm \infty$.\\
     
     Here the symmetry is perfect between the future and the past, and the Quantum Fidelity
     has no recurrences to 1, but instead decays exponentially fast to 5/9.
     
     \bigskip

     \includegraphics{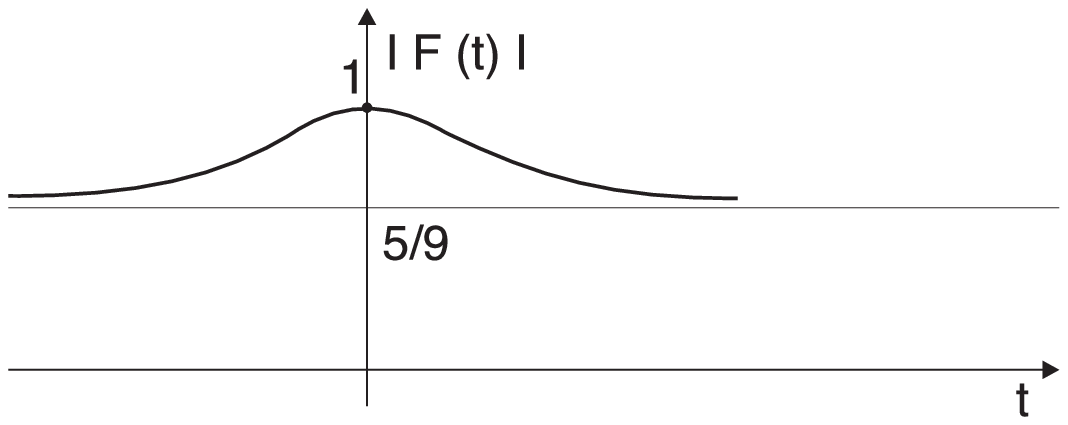}
     
     \section{CONCLUSION}
  It has been suggested in many recent works that the type of {\bf decay} of the
  Quantum Fidelity may help to {\bf discriminate} between chaotic or regular underlying classical
  motion; in other words the hypersensitivity of quantum dynamics under small perturbations,
  as measured by the type of decrease of the Quantum Fidelity, could be a signature of what
  is often called (rather improperly) ``Quantum Chaos''.\\
  
  Thus it is highly desirable to have a better understanding of how this function of time 
  (represented by equation (2) for suitable class of reference quantum states $\psi$)
   behaves at infinity in general as well in particular systems.
  In this paper we have been able to describe the full time-behavior of the Quantum Fidelity
  for a rather specific class of systems, and for reference states in a suitable class of ``generalized
  coherent states. The underlying $SU(1,1)$ structure possessed by these systems allows us to
  perform an exact calculus of the Quantum Fidelity, and to compare it with the ``Classical
  Fidelity'' of the corresponding classical motion. This classical motion can be either stable, or
  unstable with positive Lyapunov exponent. The Quantum Fidelity has the following
  remarkable properties:\\
  
  -either it decays to a (generally non-zero) limit in the past and in the future\\
  -or it manifests an infinite sequence of recurrences to 1 as time evolves.\\
  
  This sheds a new light on this question which has been addressed in a great variety of cases
  in the physics literature, and where the Quantum Fidelity is generally claimed to decay very rapidly
  to zero. Thus the first mathematical study presented here on the long time behavior of the
  Quantum Fidelity could allow in the future a better understanding of these features in more
  general situations on a rigorous level.
  
  \bigskip
 {\bf Acknowledgements} The author thanks F. Gieres and M. Kibler for interesting
      discussions about this topic. She is also very indebted to Boris Mityagin for important
      information about solutions of Hill's equation (see ref \cite{mi}).
      
      \bibliographystyle{amsalpha}

\end{document}